\documentclass[useAMS,usenatbib]{mn2e}
\usepackage{graphicx}
\usepackage[usenames]{color}
\usepackage[normalem]{ulem}

\title[M dwarf search for pulsations within Kepler GO program]{M dwarf search for pulsations within Kepler GO program}

\author[C. Rodr\'\i guez-L\'opez, J. E. Gizis, J. MacDonald, P. J. Amado and A. Carosso]{C. Rodr\'\i guez-L\'opez$^{1}$\thanks{E-mail:
crl@iaa.es (CRL); gizis@udel.edu (JEG); jimmacd@udel.edu (JM); pja@iaa.es (PJA); carosso@udel.edu (AC)}, J. E. Gizis$^{2}$, J. MacDonald$^{2}$, P. J. Amado$^{1}$ and A. Carosso$^{2}$\\
$^{1}$Dep. de F\'\i sica Estelar. Instituto de Astrof\'\i sica de Andaluc\'\i a (IAA-CSIC), 18008 Granada, Spain\\
$^{2}$Dept. of Physics and Astronomy, University of Delaware, Newark, DE 19716, USA }

\begin{document}

\date{Accepted 2014 Sept 04. Received 2014 Sept 04; in original form 2014 Sept 04}

\pagerange{\pageref{firstpage}--\pageref{lastpage}} \pubyear{2014}

\maketitle

\label{firstpage}

\begin{abstract}
We present the analysis of four M dwarf stars -plus one M giant that seeped past our selection criteria- observed in Cycle 3 of Kepler Guest Observer program (GO3) in a search for intrinsic pulsations. Stellar oscillations in M dwarfs were theoretically predicted by \cite{crl12} to be in the range $\sim$20-40~min and $\sim$4-8~h, depending on the age and the excitation mechanism. We requested Kepler short cadence observations to have an adequate sampling of the oscillations. The targets were chosen on the basis of detectable rotation in the initial Kepler results, biasing towards youth.The analysis reveals no oscillations attributable to pulsations at a detection limit of several parts per million, showing that either the driving mechanisms are not efficient in developing the oscillations to observable amplitudes, or that if pulsations are driven, the amplitudes are very low. The size of the sample, and the possibility that the instability strip is not pure, allowing the coexistence of pulsators and non-pulsators, prevent us from deriving definite conclusions. Inmediate plans include more M dwarfs photometric observations of similar precision with Kepler K2 mission and spectroscopic searches already underway within the \emph{Cool Tiny Beats Project} \citep{anglada2014, berdinas2014}  with the high-resolution spectrographs HARPS and HARPS-N.

\end{abstract}


\section{Introduction}
  Asteroseismology has already proved to be able to determine the radius, the mass and the age of a Solar-like star to about 3, 5 and 5-10 per cent uncertainties, respectively (see \citealt{moya13} and references therein),with the precision in the results being a direct consequence of how well the physical processes are implemented in the theoretical models.

\cite{baran11a} first mentioned that theoretical calculations predicted the instability of M dwarf models driven by the $\epsilon$ mechanism of the He$^3$ burning. A detailed theoretical study by \cite{crl12} predicted the instability of the fundamental radial mode in solar metallicity M dwarf models. Two different thermodynamic methods work to excite the oscillations: (1) An $\epsilon$ mechanism associated with deuterium burning produces the excitation of young low-mass models with periods in the range of about 4 to 8~hours; and the same mechanism linked to the He$^3$ burning excites old low-mass models in the range of about 20 to 30~minutes. All the models excited by this $\epsilon$ mechanism are completely convective (0.10 to 0.25~M$\odot$). (2) Periodic blocking of the radiative flux at the tachocline, known as the 'flux-blocking' mechanism for models older than 500~Myr that are not fully convective (0.30 to 0.60~M$\odot$) yields excited periods in the range of about 35-40~minutes. New calculations by \cite{crl14} for a wider M dwarf model grid broaden the instability region to from 20~min to 3~h for main sequence M dwarfs and from 4 to 11~h for young M dwarfs, and also extends the instability to non-radial and non-fundamental modes.

  A ground-based search for M dwarf pulsations (\citealt{baran11b,krzesinski12,baran13a}) observed a total of 120 M0-M4 M dwarfs with an overall 1~mmag precision, with no detections. \cite{baran11a} analysed the light curves of 86 Kepler \citep{borucki10} stars preliminary tagged as M dwarfs, from which only six survived as M dwarfs after spectroscopic observation. Those six M dwarfs had short cadence (58.85~s sampling) light curves from Q2 or Q3 Kepler public data releases. None was confirmed to be variable above a 1-10~$\mu$mag threshold. Together the Kepler and ground-based results indicate that if pulsations occur in M dwarfs, their amplitudes are very low and they are difficult to detect. 

 The observational discovery of pulsations in M dwarfs is hampered mainly by flares, spots and atmospheric activity, as well as by the intrinsic faintness of the objects and by ignorance of the amplitudes of the oscillation, which cannot be predicted by the usual linear oscillation codes.

  We explored the ESO public archive\footnote{http://archive.eso.org/eso/eso$\_$archive$\_$main.html} for radial velocity data of nearby M dwarfs appropriate to search for pulsations. Most of the data come from the HARPS spectrograph exoplanets search \citep{bonfils11}. However, these data were not useful for our purposes due to a time sampling much longer than needed to be able to detect signals with periods as short as 20~minutes. The general trend in the publicly available photometric or radial velocity data of M dwarfs is that they have a sparse sampling which would conceal the short periodic signals that we are looking for.

  To overcome this problem we applied for Short Cadence observations in Cycle 3 of the Kepler Guest Observer (GO) program in December 2010, before the papers of Baran et al. were published. Kepler data give the best possible photometric precision currently achievable, with detection limits of only a few parts per million (ppm) for M dwarf stars.

  Our target selection is described in Section~2. The data analysis is presented in Section~3 and our conclusions are given in Section~4.

\section{Target selection}

 To cover the shortest oscillation periods expected, we requested Short Cadence (SC) observations of five selected M dwarfs in Cycle 3 of the Kepler Guest Observer program (GO3) (proposal GO30021). Target pixel files were made available, as well as two types of processed data: (1) Simple Aperture Photometry (SAP) data, minimally processed, corrected for cosmic rays and background removed, and (2) Pre-search Data Conditioning (PDCSAP) data in which systematic correction to the light curves was done prior to a transit search, and which should be used with caution for other astrophysical analysis, such as asteroseismology.

\begin{table}
\caption{M dwarf targets}
\label{tab:targets}
 \begin{minipage}{50mm}
\renewcommand{\footnoterule}{}
\renewcommand{\tabcolsep}{4pt}
\begin{tabular}{ccccccc}
\hline
  KIC    &  Other      & Kep     & 4$\sigma$ & J-H   &  H-K  &   SC Qn   \\
  ID     &    ID       & mag     &  (ppm)    &       &       & (months)  \\
\hline
 002424191 &           & 11.880  &  6.6      & 0.741 & 0.239 &  Q11 (3)  \\
\hline
 004142913 &  GJ\,4099 & 10.910  &  2.5      & 0.576 & 0.228 & Q8\footnote{public data} (2)    \\
           &           &         &           &       &       & Q10 (3)   \\
\hline
 004743351 &           & 13.806  &  16.2     & 0.631 & 0.174 & Q13 (3)   \\
\hline
 008607728 &           & 13.410  &  12.2     & 0.554 & 0.229 & Q12 (3)   \\

\hline
 009726699 &  GJ\,1243 & 12.738  &  15.7     & 0.541 & 0.272 &  Q6$^{a}$  (2)  \\
           &           &         &           &       &       &  Q10 (3)  \\
           &           &         &           &       &       &  Q12 (3)  \\
           &           &         &           &       &       &  Q13 (3)  \\
\hline
\end{tabular}
\end{minipage}
\end{table}

  We selected five targets with detected variability and rotation periods from \cite{basri11}'s analysis of the Kepler Q1 dataset, which biases our sample towards younger ages. Three were previously known nearby M dwarfs (KIC\,004142913, KIC\,008607728, and KIC\,009726699) and two were newly identified M dwarfs (KIC\,002424191 and KIC\,004743351).  

  KIC\,004142913 (GJ\,4099 or StKM\,1-1680) was discovered by \cite{Stephenson1986} in an objective prism survey. \cite{reid1995} classified it as M1V and estimated the distance to be $\sim$20~pc. It has H$\alpha$ absorption indicating the presence of a weak chromosphere \citep{gizis2002}. It has a $\sim$35~day rotation period with a significant spot. KIC\,008607728 (LP\,230-6), discovered by \cite{Luyten1979a, Luyten1979b}, has been little studied. The $V$ magnitude of 13.86 and $V-J=3.54$ \citep{lepshara2005} suggests a distance $\sim$50~pc according to the main sequence relationship of \cite{lepine2005}, and the color is consistent with spectral type $\sim$M1V. A clear spot signature indicates a rotation period of $\sim$30~days. KIC\,009726699 (GJ\,1243, G\,208-042) was discovered as a proper motion star \citep{giclas1971}. \cite{Harrington1980} measured a trigonometric parallax of $84.6 \pm 2.4$~mas (i.e., $11.8 \pm 0.4$~pc), so it is one of the nearest stars in the Kepler field. It was classified as M4Ve by \cite{reid1995}. The H$\alpha$ emission is due to the rapid rotation of the star: \cite{Reiners2012} measured $v \sin i =$ 22~km s$^{-1}$ and noted $i \approx 83^\circ$ given the rotation period of 0.59 days seen in both ground-based \citep{irwin11} and Kepler photometry. \cite{savanov11} have modeled the complex starspot coverage in early Kepler long-cadence data. KIC\,004743351 is the second brightest of \cite{basri11} new M dwarfs with a period of $\sim$10~days. Both the KIC and the 2MASS colors indicate dwarf status. A small proper motion is revealed on the finder charts. \cite{Dressing2013} fit models to photometry to estimate that it has T$_{eff} = 3918^{+74}_{-79}$~K, M$= 0.536^{+0.06}_{-0.05} M_\odot$, and a distance of $130^{+19}_{-14}$~pc. No spectral type has been published for this star, but comparison to the \cite{Muirhead2012}'s table of effective temperatures and spectral types for Kepler Objects of Interest indicates that it should be an M0V. Finally, we chose KIC\,002424191 despite its lack of apparent motion because it was one of the brightest objects classified as a dwarf in the \cite{basri11} analysis. However, \cite{Mann2012} have observed it spectroscopically and classified it as an M giant; we note also that it has no flares. We analysed KIC\,002424191 anyway, as it formed part of our GO proposal.

  We requested to observe four of the targets for a quarter and  KIC\,009726699 for the full year, as its flare activity would require a long time baseline to confirm oscillations. In addition, the SC data of a flare star would yield a very good time resolution of the flares, which is potentially very valuable to study the fine structure of the flares and to contribute to their statistical characterization, such as their frequency or energy distribution \citep{hilton10}.

   \begin{figure*}
     \begin{tabular}{cc}
 \resizebox{0.48\linewidth}{6cm}{\includegraphics{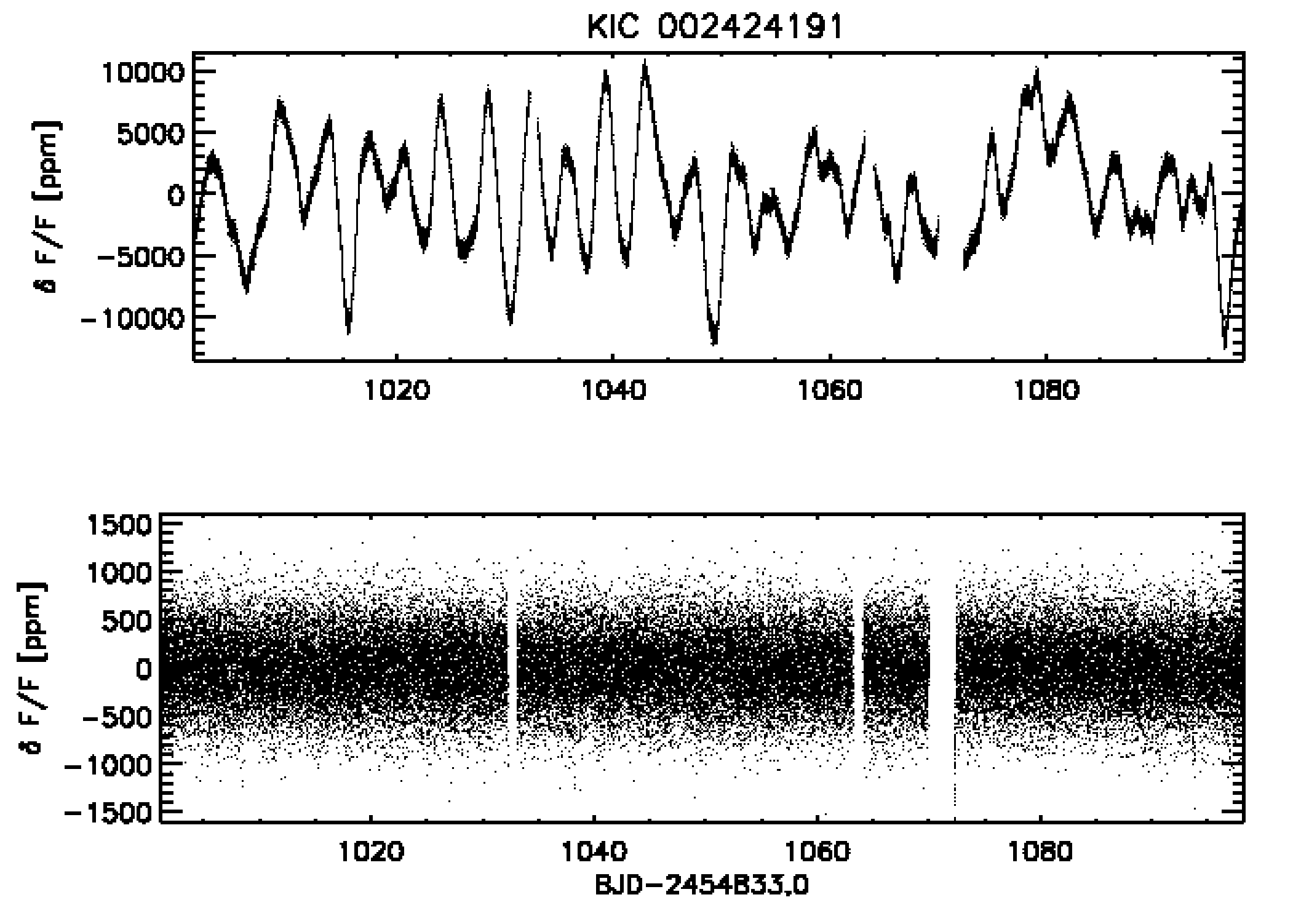}} &
   \resizebox{0.48\linewidth}{6cm}{\includegraphics{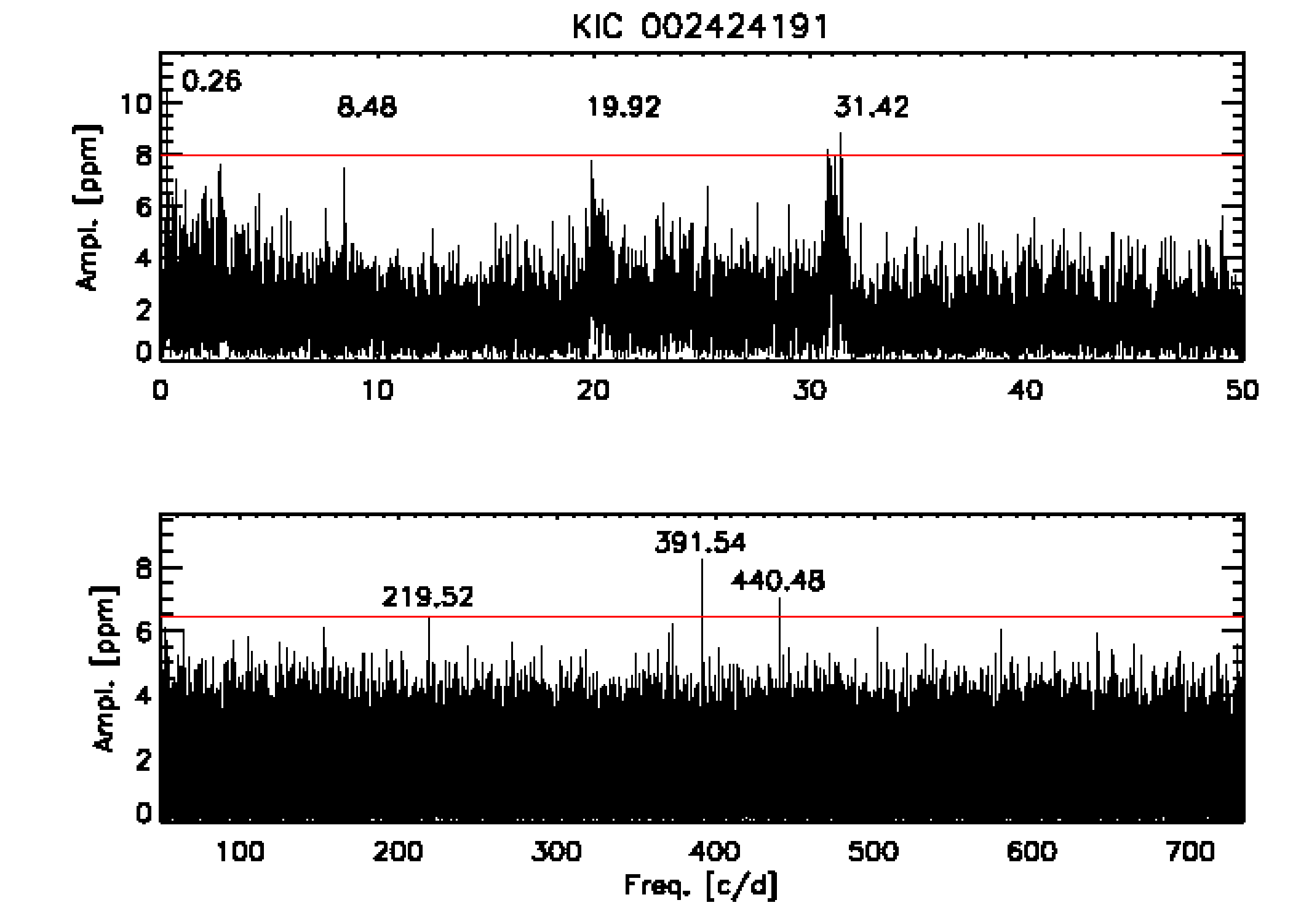}} \\
      \end{tabular}
   \caption{Left: KIC\,002424191 light curve in Q11. Above: corrected for outliers with a 3$\sigma$ clipping of the two-point difference function, converted to relative flux in ppm and fitted a 1-degree polynomial. Below: a binning of 5~h, 3~h in the last month of the quarter, was used to perform a cubic spline interpolation and remove the quasi-periodic variations due to stellar activity. Right: Amplitude spectrum and prominent frequencies in the low frequency region (above) and up to the Nyquist frequency (below). The solid red line is four times the mean amplitude in the plotted frequency range, which is higher in the low frequency region. See the text for details.}
   \label{fig:2424191}
   \end{figure*}

  Table~\ref{tab:targets} shows the KIC identification of each target, its Gliese \citep{gliese95} identification if any, its Kepler magnitude, a threshold detection limit for all available data on the target, calculated as four times the mean amplitude in the whole frequency range up to the Nyquist frequency, the $J-H$ and $H-K$ MAST colour indices, and the quarters in which the target was observed. We note that, as well as our proprietary data from GO3, corresponding to quarters Q10 to Q13 (made available from January to October 2013, respectively), we also analysed public SC data available for KIC\,004142913 and KIC\,009726699, as indicated in the table. The colour indexes $J-H$ and $H-K$ of all the targets satisfy the colour cut $J-H < 0.75$, $H-K > 0.1$ used to separate giants from dwarfs (see \citealt{ciardi11} figure~4).

\section{Data Analysis}
  We preferred to use the SAP non-processed raw data instead of the PDCSAP data, in case some astrophysical signal of our interest has been removed from the PDCSAP processed data, and implemented our own corrections to the light curves. All data were 3-sigma clipped to a point-to-point deviation of the two point different function to remove outliers, following \cite{garcia11}.  

  The SAP flux in e/s was converted to relative flux in parts per million (1~ppm $\simeq$ 1.086~$\mu$mag) and a 1-degree polynomial was fitted to remove linear trends. The light curves were then appropriately binned, between 3 to 12~h depending on the target, before performing a cubic spline interpolation to detrend the original light curve to remove the variations due to spots and obtain a zero-centered flux. No attempt was done to correct for small jumps that changed the mean value of the light curve, or small temperature drifts as they were better corrected with the spline.

  Period04 \citep{lenz05} was used to perform a Fourier Transform (FT) analysis for each of the targets. Each month of a quarter was analysed separately and then the three months of the quarter were merged together. No flux correction was necessary in the merging, as fluxes were already zero-centered in the previous step of the analysis.-

  Fourier Transforms were calculated from 0 up to the Nyquist frequency (corresponding to $\sim$~2~min) in search for very short period oscillations. This would be in the range of expected periods for stochastic oscillations if solar-like scaling relations apply: following \cite{chaplin11}, we have estimated the frequency of maximum power for an arbitrary M star with fundamental radial stochastically excited modes: 

\begin{equation}
 \nu_{\rmn{max}}=\nu_{\rmn{max,\odot}} \left({{M} \over {M\odot}}\right)  \left({{R} \over {R\odot}}\right)^{-2}  \left({{T_{\rmn{eff}}} \over {T_{\rmn{eff},\odot}}}\right)^{-0.5}
\end{equation}

  where $\nu_{\rmn{max},\odot}=$3150~$\mu$Hz and $T_{\rmn{eff},\odot}=$5777~K. 
  Adopting representative values for mass, radius and $T_{\rmn{eff}}$ for M dwarf models given in table~1 of \cite{crl12}, the period corresponding to the frequency of maximum power is of the order of 1-2~min for the 0.25 to 0.50M$\odot$ models. A more thorough discussion on possible stochastic excitation of modes for M dwarfs is given in \cite{crl14}, where again theoretical evidence for periods around 2~min is given for M dwarf models on the main sequence.

  Frequencies were considered significant if their amplitude signal to noise ratio calculated in a box of 2~c/d around the peak under consideration was larger than 4 \citep{breger93}, which corresponds to a 99.9\% confidence level \citep{kuschnig1997, reegen2004} and which is the criterium usually accepted for significance. 

   \begin{figure*}
     \begin{tabular}{cc}
\resizebox{0.48\linewidth}{6cm}{\includegraphics{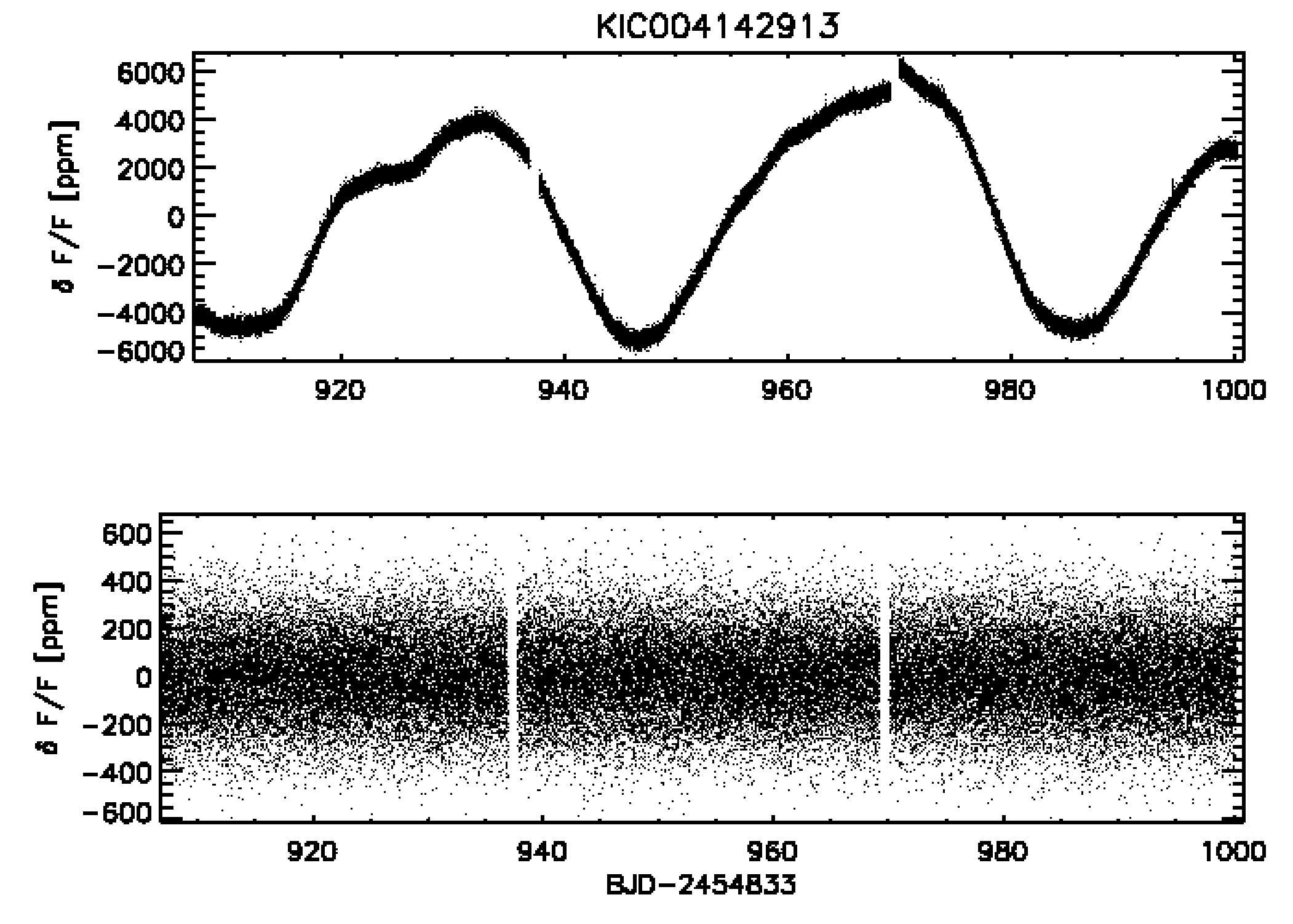}} &
   \resizebox{0.48\linewidth}{6cm}{\includegraphics{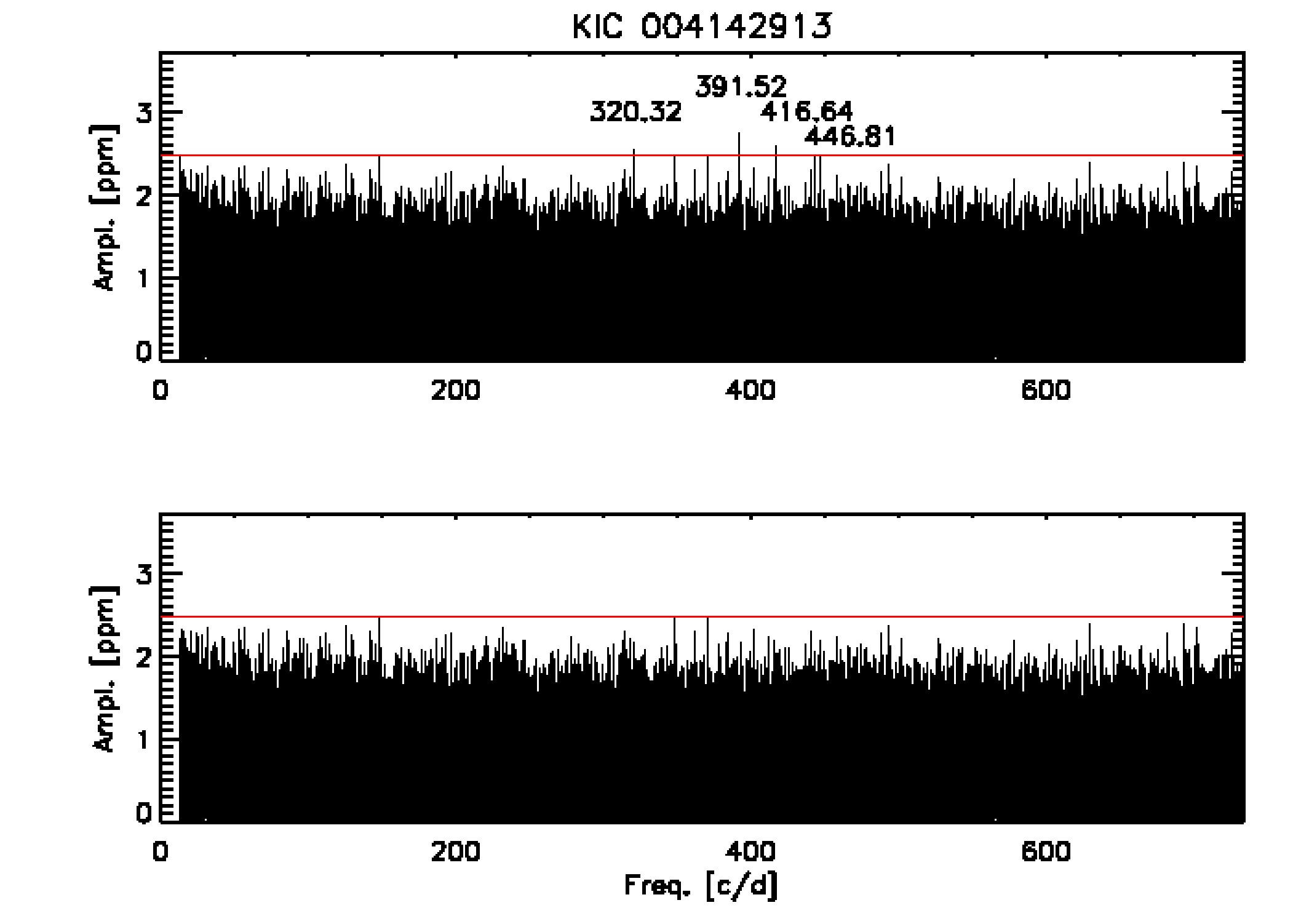}}
      \end{tabular}
   \caption{Left: KIC\,004142913 light curve in Q10. Top: corrected for outliers with a 3$\sigma$ clipping of the two-point difference function and converted to relative flux in ppm. Below: a binning of 4~h was used to perform a cubic spline interpolation and remove the rotational modulation and activity. Right: (Above) Amplitude spectrum and prominent frequencies. (Below) Amplitude spectrum after pre-whitening of six frequencies. See the text for details.}
   \label{fig:4142913}
   \end{figure*}

  Every frequency found with the Fourier transform analysis was checked against the known spurious frequencies listed in the Kepler Data Characteristic Handbook\footnote{http://archive.stsci.edu/kepler/manuals/Data\_Characteristics.pdf} (DCH) and the Kepler Data Release Notes\footnote{http://archive.stsci.edu/kepler/data\_release.html} for each quarter. In addition, it was carefully matched to the list of artifacts and frequencies throughly described in the helpful work of \cite{baran13b}.

  We note here that we found spurious frequencies corresponding to the 6th, 7th, 8th and 9th harmonic of the long cadence readout in several of our targets, as well as examples of the wide artifacts between 20 and 35~c/d ('wide 20+'), of the 35.7~c/d comb (U artifact), and possibly of the single peak W, all described in \cite{baran13b}.

\subsection{KIC\,002424191}
  KIC\,002424191 was observed in Q11. The Kepler SAP light curve corrected for outliers, converted to relative flux in ppm and 1-degree polynomial fitted, as well as the spline detrended light curve are shown in Fig.~\ref{fig:2424191} (left). The first two gaps in the light curve correspond to monthly downloading of data to Earth, while the last one is due to a safe mode event occurring for $\sim$2.5~days (starting on December 7th 2011). The amplitude spectrum is shown in Fig.~\ref{fig:2424191} (right), where the most prominent frequencies, that will be discussed below, were labeled. It was split into two plots: from 0 to 50~c/d and from 50 to 733~c/d (Nyquist frequency) to show the low and high frequency range, respectively. The solid red line is four times the noise level, calculated as the mean amplitude in the plotted frequency range to give a global vision of which frequencies may be significant. The mean noise level is usually higher in the low frequency range, due to long term variations in the light curve that are not completely removed in the data processing, or to frequencies having more amplitude in that region. In Table~\ref{tab:targets} we give a detection threshold, 4$\sigma$, calculated as four times the mean noise level up to the Nyquist frequency for all the available data, to give a guide to the precision level attained for each target. The significance criterion will be more restrictive, as it will evaluate the noise level in a box of 2~c/d around the frequency under consideration.

  The Fourier analysis of Q11 quarter found significant the spurious frequencies marked in Fig.~\ref{fig:2424191} (lower right): 391.54~c/d (3.7~min, 8/LC, 8.2~ppm) and 440.48~c/d (3.3~min, 9/LC, 7.0~ppm), namely the 8th and 9th harmonic of the LC sampling. Barely reaching the significant limit is 219.52~c/d (6.6~min, 6.4~ppm). Also significant was found a frequency with an amplitude of 9~ppm at 31.42~c/d Fig.~\ref{fig:2424191} (upper right) with a corresponding period of 45~min. This period falls within the range of excited ones predicted in \cite{crl12} and \cite{crl14}. As exciting as this may seem, this frequency can be matched to the spurious frequency 31.35~c/d noted as very broad in the Kepler DCH, within the range 31.17-31.53~c/d and also well documented as a spurious frequency of the Q11 quarter in \cite{baran13b}, denoted the 'wide 20+' artifact. Indeed, this frequency presents a 1~c/d wide structure, and frequencies within this range are subsequently found in the amplitude spectrum, slightly in the 4$\sigma$ detection limit. When the periodogram is calculated separately for each month of the quarter, the artifact changes its peak frequency from about 20 to 31~c/d in exactly the same way and with the same structure as those in fig.~16 in \cite{baran13b}.

  Some prominent non-significant frequencies marked in Fig.~\ref{fig:2424191} (upper right) were: a low frequency at 0.26~c/d (3.8~d) associated with the non complete removal of long trends in the light curve; 8.48~c/d (2.8~h), half of the listed spurious 16.98~c/d frequency and finally 19.92~c/d (1.2~h), that can be matched to the 20.95~c/d artifact listed in the Kepler spurious frequency list and also to the 'wide 20+' artifact of the Q11.1 in \cite{baran13b}.

  \begin{figure*}
     \begin{tabular}{cc}
\resizebox{0.48\linewidth}{6cm}{\includegraphics{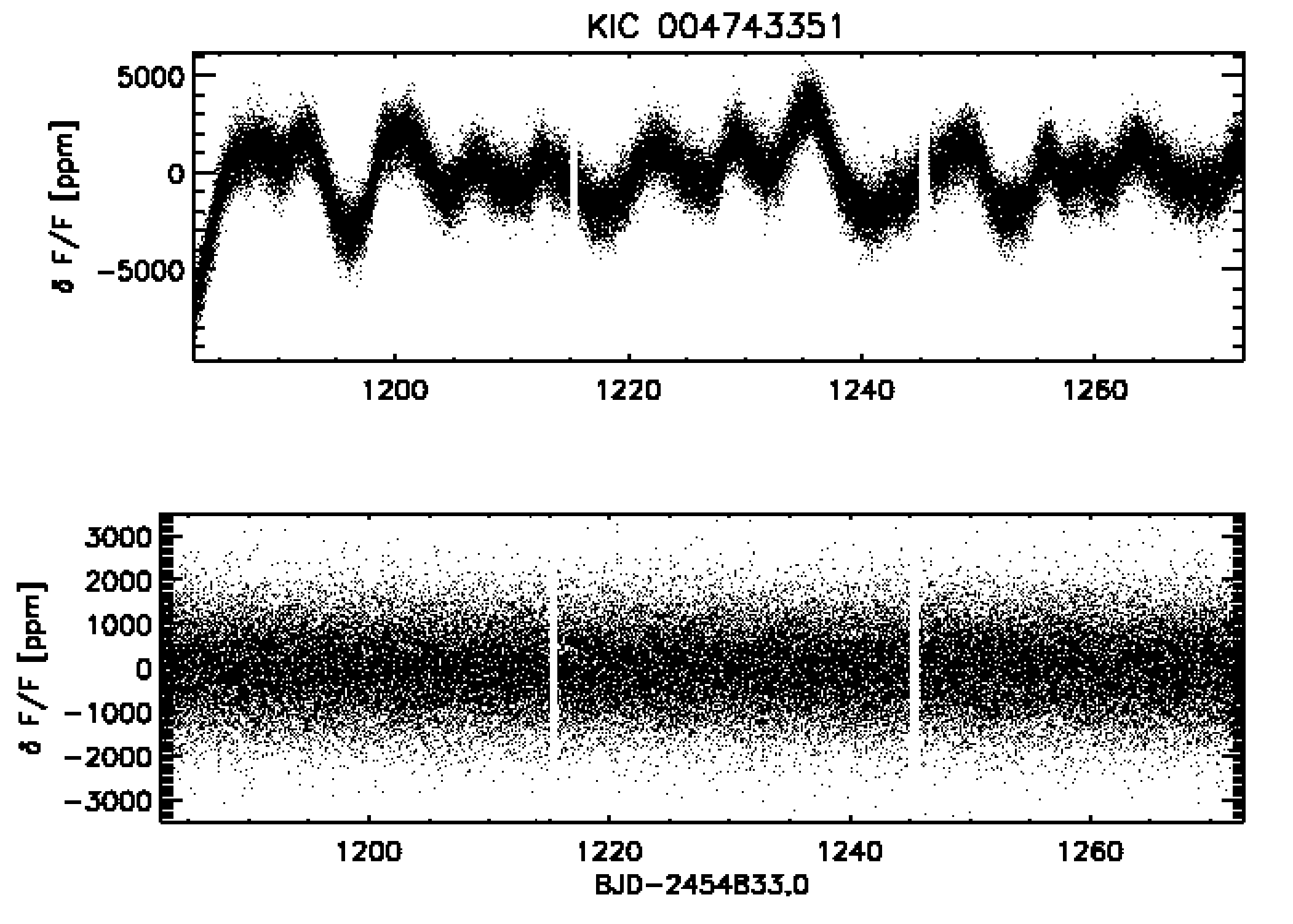}} &
   \resizebox{0.48\linewidth}{6cm}{\includegraphics{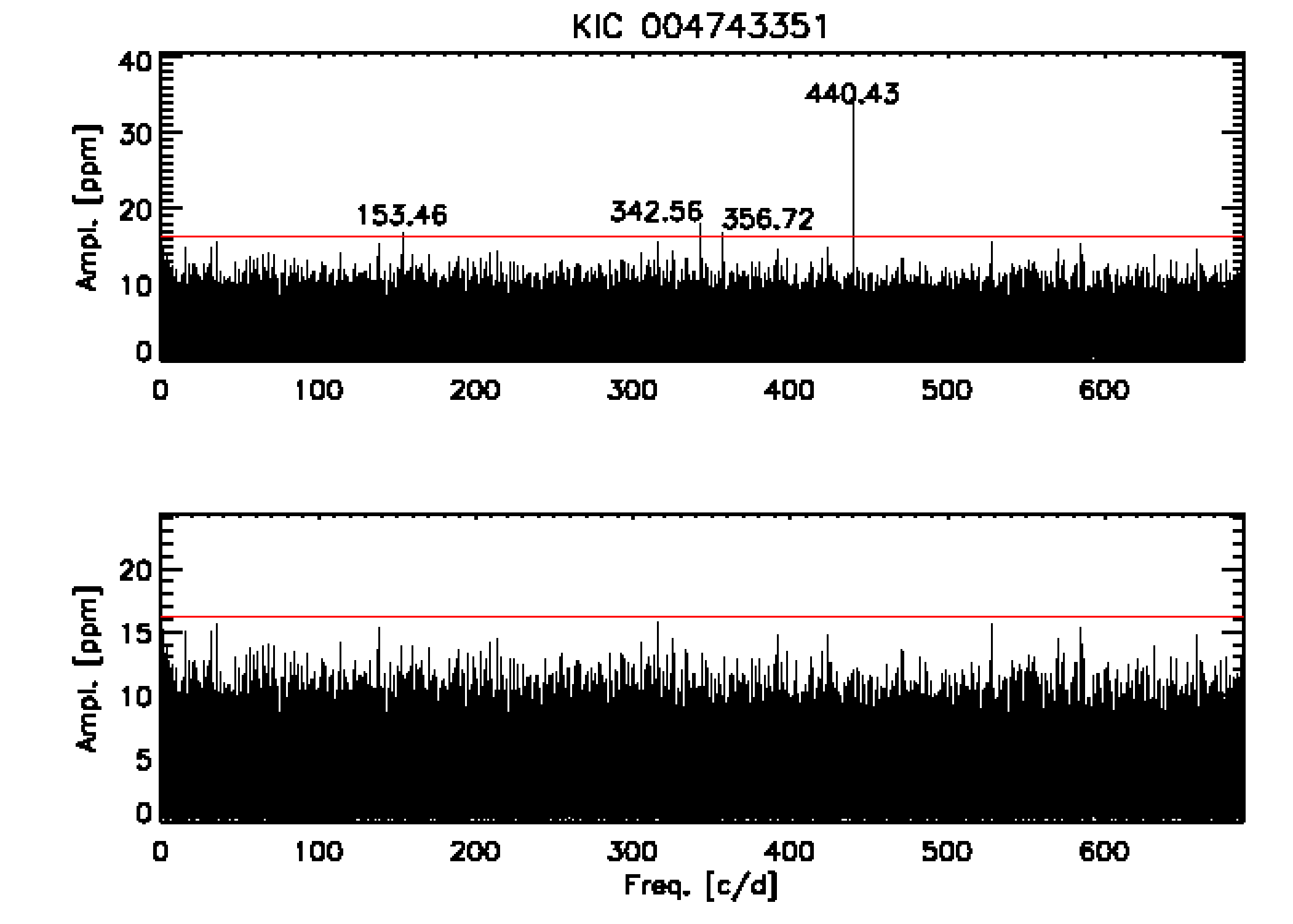}}
      \end{tabular}
   \caption{Left: KIC\,0047433511 light curve in Q13. Above: corrected for outliers with a 3$\sigma$ clipping of the two-point difference function, converted to relative flux in ppm and detrended with a 1-degree polynomial. Below: a binning of 12~h was used to perform a cubic spline interpolation and remove the quasi-periodic variations due to stellar activity. Right: (Above) Amplitude spectrum and prominent frequencies. (Below) Amplitude spectrum after pre-whitening of six frequencies. See the text for details.}
   \label{fig:4743351}
   \end{figure*}

\subsection{KIC\,004142913}
 KIC\,004142913 was observed in Q10 and SC public data of the first two months of Q8 were added to the analysis. The Kepler SAP light curve corrected for outliers and converted to relative flux in ppm, as well as a 4~h spline fitted light curve are shown in Fig.~\ref{fig:4142913} only for Q10 for clarity reasons. The light curve was not 1-degree polynomial fitted due to its evident rotational modulation. The gaps in Q10 are due to monthly data downloads, while a safe mode event occurred in Q8.2 and the corresponding data were lost. A crude FT of the light curve without any spline fitting yields a rotation period of 35~days.  This is the brightest object of our sample, so the mean noise of the light curve is substantially lower than for the other objects, reaching an impresive 0.6~ppm, which gives a mean detection threshold of 2.5~ppm. The amplitude spectrum is shown in Fig.~\ref{fig:4142913} (right, above), where the most prominent frequencies, that will be discussed below, were labeled; and in Fig.~\ref{fig:4142913} (lower right), after pre-whitening of six frequencies.

 We performed the FT analysis of the Q8.1, Q8.2 and Q10 combined, searching for frequencies starting at 12~c/d to remove low frequencies as a result of an imperfect removal of the variations caused by the activity of the star. We found 391.52~c/d (8/LC, 2.7~ppm) significant, and barely above the 4$\sigma$ limit a bunch of frequencies in the 3 to 5 minute range: 416.64~c/d (3.5~min, 2.6~ppm), 320.32~c/d (4.5~min, 2.6~ppm), 446.81~c/d (3.2~min, 2.5~ppm) and 442.58~c/d (3.3~min, 2.5~ppm), all of them marked in Fig.~\ref{fig:4142913} (lower right) except the last one for clarity reasons, and a frequency at 13.22~c/d considered due to incomplete removal of activity signals in the light curve.

 From the analysis of the Q8 months together no frequency was found significant, while from the analysis of the Q10 data we found:  391.52~c/d (8/LC) and 320.32~c/d, but none of the others in the 3 to 5 minute period range. While this could be attributed to the noise level achieved for each analysis (Q8+Q10 $\sigma$=0.6~ppm, Q10 $\sigma$=0.8~ppm, Q8 $\sigma$=1.1~ppm), we found the following: the 320.32~c/d frequency is only present in one of the Q8 months and in all individual Q10 months, always not significantly and changing its amplitude from month to month. Both 416.64~c/d and 446.81~c/d frequencies are not present in one of the Q10 months and change their amplitude the rest of the individual months, not being significant in any of them. Also the 442.58~c/d frequency changes its amplitude from month to month, not reaching enough significance and not even having the largest amplitude in the considered 2~c/d box. As a result, we conclude that we can not identify any of these frequencies with a stable pulsation frequency.

\subsection{KIC\,004743351}
 KIC\,004743351 was observed in Q13 and it is the faintest object of our sample. The Kepler SAP light curve corrected for outliers, converted to relative flux in ppm and 1-degree polynomial fitted, as well as the spline detrended light curve are shown in Fig.~\ref{fig:4743351}. The two gaps in the curve are due to the spacecraft monthly download of data to Earth. The amplitude spectrum is shown in Fig.~\ref{fig:4743351} (upper right) and after pre-whitening of the marked frequencies (lower right). This object is the faintest of our sample and consequently has the highest detection threshold, 16.2~ppm.

  The analysis of the whole Q13 data revealed the harmonics of the LC sampling: 440.43~c/d (9/LC, 33.8~ppm) and 342.56~c/d (7/LC, 18.2~ppm) and two frequencies: 153.46~c/d (9.4~min, 17.0~ppm)  and 356.72~c/d (4.0~min, 16.9~ppm) just above the the 4$\sigma$ limit.

  The peak at 153.46~c/d is present in the analysis of two out of the three individual months. Each month has a similar mean noise of about 7~ppm in a 2~c/d box around the considered peak, but the peak is significant in only one of those two months, and in the other is not even the peak with highest amplitude in the considered box. Moreover, it is not well pre-whitened in the amplitude spectrum of the whole quarter, as it appears again as 153.42~c/d. Both facts point at stellar activity, although it may also be the 'W artifact' noted in \cite{baran13b} as a single frequency usually around 155-158~c/d.

  A peak at about 356.72~c/d is present insignificantly in the analysis of the individual months (mean noise in a box of 2~c/d around the peak goes from 6.3 to 7.1~ppm), not even being the peak with the highest amplitude within the considered box and changing amplitude among the different months. The peak becomes significant in the analysis of the whole quarter (lower noise), although it is not well pre-whitened, appearing again with a modified phase. All this, together with the fact that in the 200-450~c/d region peaks at S/N=4 are frequent in Kepler data of other stars (Baran, personal comm.), we can not conclude that this frequency is attributable to pulsations and more data would be desirable to reach a definite conclusion.

  From the analysis of the individual months of the quarter, in Q13.1 we recover a strong peak at 31.49~c/d with the same structure as the artifact in fig.18 of \cite{baran13b}.

\subsection{KIC\,008607728}

  \begin{figure*}
     \begin{tabular}{cc}
 \resizebox{0.48\linewidth}{6cm}{\includegraphics{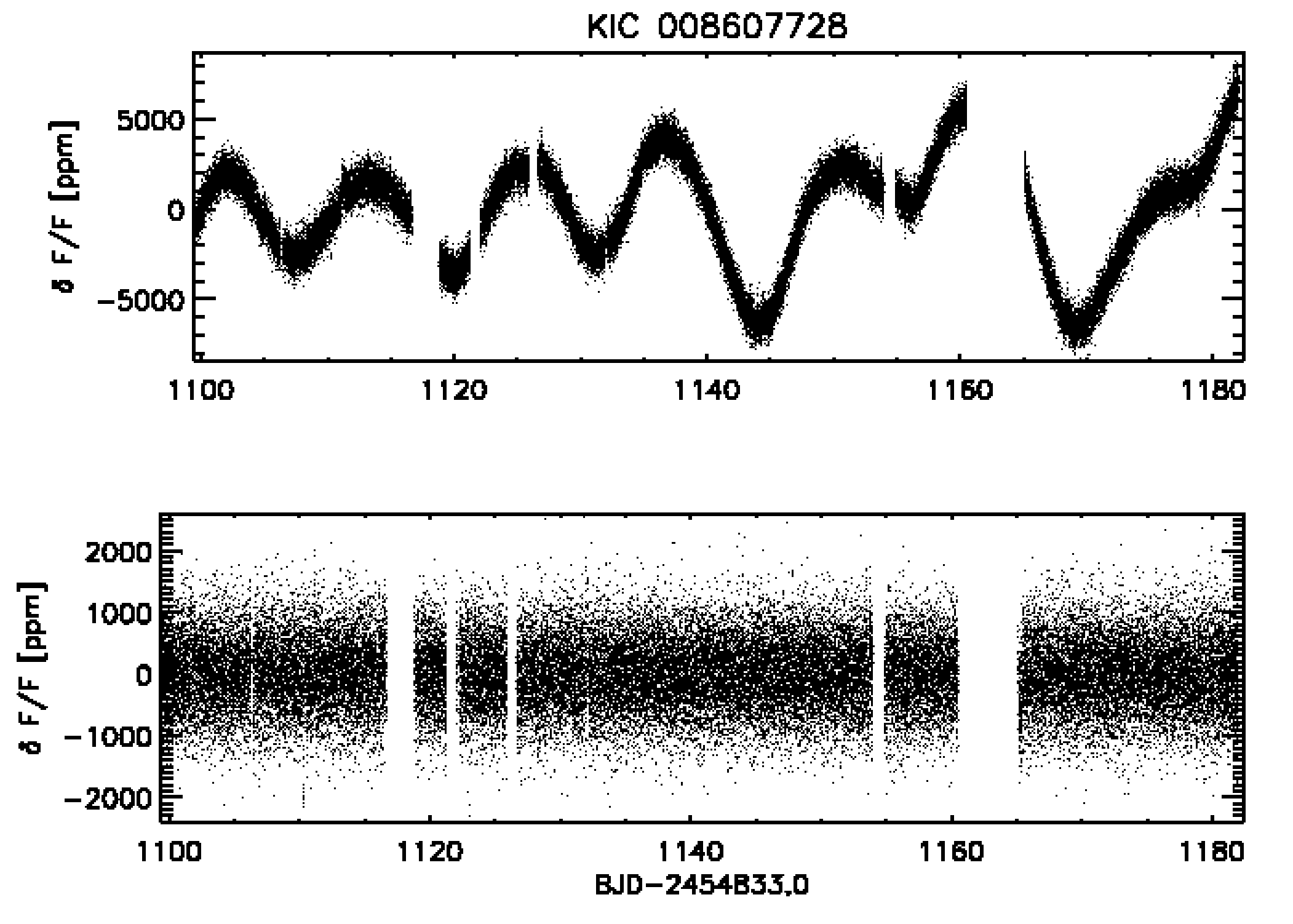}} &
   \resizebox{0.48\linewidth}{6cm}{\includegraphics{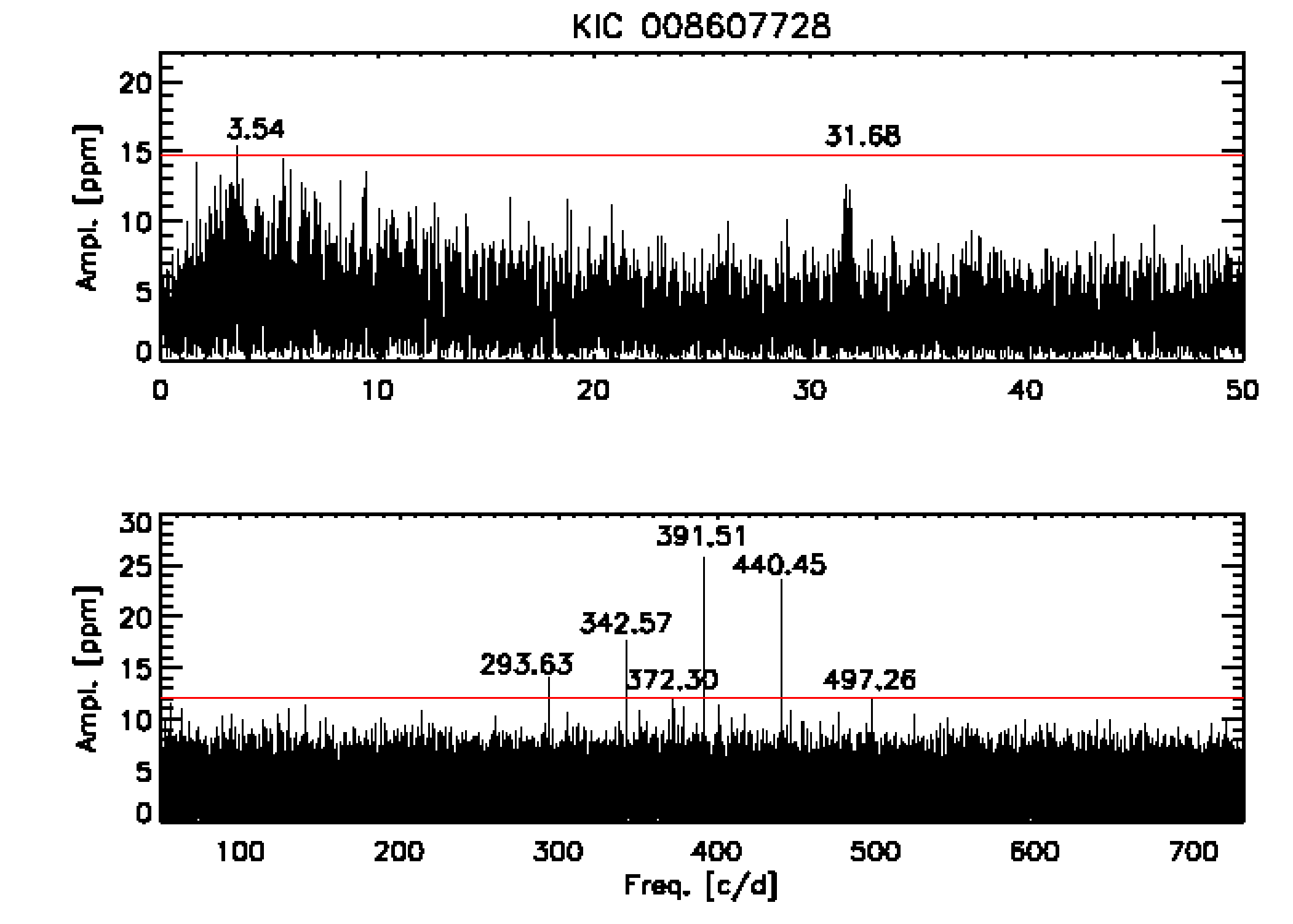}} \\
      \end{tabular}
   \caption{Left: KIC\,008607728 light curve in Q12. Above: corrected for outliers with a 3$\sigma$ clipping of the two-point difference function, converted to relative flux in ppm and 1-degree polynomial fitted. Below: a 4~h binning, 8~h in the third month of the quarter, was used to perform a cubic spline interpolation and remove the variations most likely due to spots. The first, second and fifth gap are data rendered unuseful due to coronal mass ejections. The third and fourth gap are the usual ones due to data being downloaded to Earth. Right: Amplitude spectrum and prominent frequencies in the low frequency region (above) and up to the Nyquist frequency (below). The solid red line is four times the mean amplitude in the plotted frequency range, which is higher in the low frequency region. See the text for details.}
   \label{fig:8607728}
   \end{figure*}

  KIC\,008607728 was observed in Q12. The Kepler SAP light curve corrected for outliers, converted to relative flux in ppm and 1-degree polynomial fitted, as well as the spline detrended light curve are shown in Fig.~\ref{fig:8607728}. The second, third and last gap in the data were due to coronal mass ejections, were the data were rendered unuseful (see Kepler Data Release Notes 17). The other gaps correspond to the usual monthly download of data to Earth or to flare events. The amplitude spectrum is shown in Fig.~\ref{fig:8607728} (right), in the low frequency region up to 50~c/d (upper) and up to the Nyquist frequency (lower). The solid red line is four times the mean noise in each region, the mean detection threshold being 12.2~ppm. The most prominent frequencies are marked in the plot; the Fourier analysis of the whole quarter only found significant the spurious frequencies 391.51~c/d (8/LC, 25.6~ppm), 440.45~c/d (9/LC, 23.7~ppm), 342.57 (7/LC, 17.7~ppm) and 293.63 (6/LC, 14.0~ppm), corresponding to 8th, 9th, 7th and 6th harmonics of the LC sampling time. We also found 372.30~c/d (3.9~min, 12.2~ppm) and 497.26~c/d (2.9~min, 12.2~ppm) just above the 4$\sigma$ limit. To check on these two frequencies we turned to the independent analysis of each month of the quarter.

  The peak at 372.30~c/d is present in every month of the quarter, although non significantly. The local mean noise in a 2~c/d box around the peak is about 5~ppm for each individual month, and the peak is the largest in the considered box only in the second month of the quarter. 

  The peak at 497.26~c/d is present only in two months of the quarter, where the local mean noise for every month in a 2~c/d box around the peak is about 5~ppm. 

  We consider both peaks not stable, because if they were so, each of them should appear with a similar amplitude in all the three individual months, as they have comparable mean noise; therefore we conclude that they are not attributable to pulsations.

  A frequency at 31.68~c/d with amplitude 12.7~ppm also found significant was identified with the 'wide 20+' artifact listed in \cite{baran13b}.

\subsection{KIC\,009726699}

  \begin{figure*}
     \begin{tabular}{cc}
 \resizebox{0.48\linewidth}{6cm}{\includegraphics{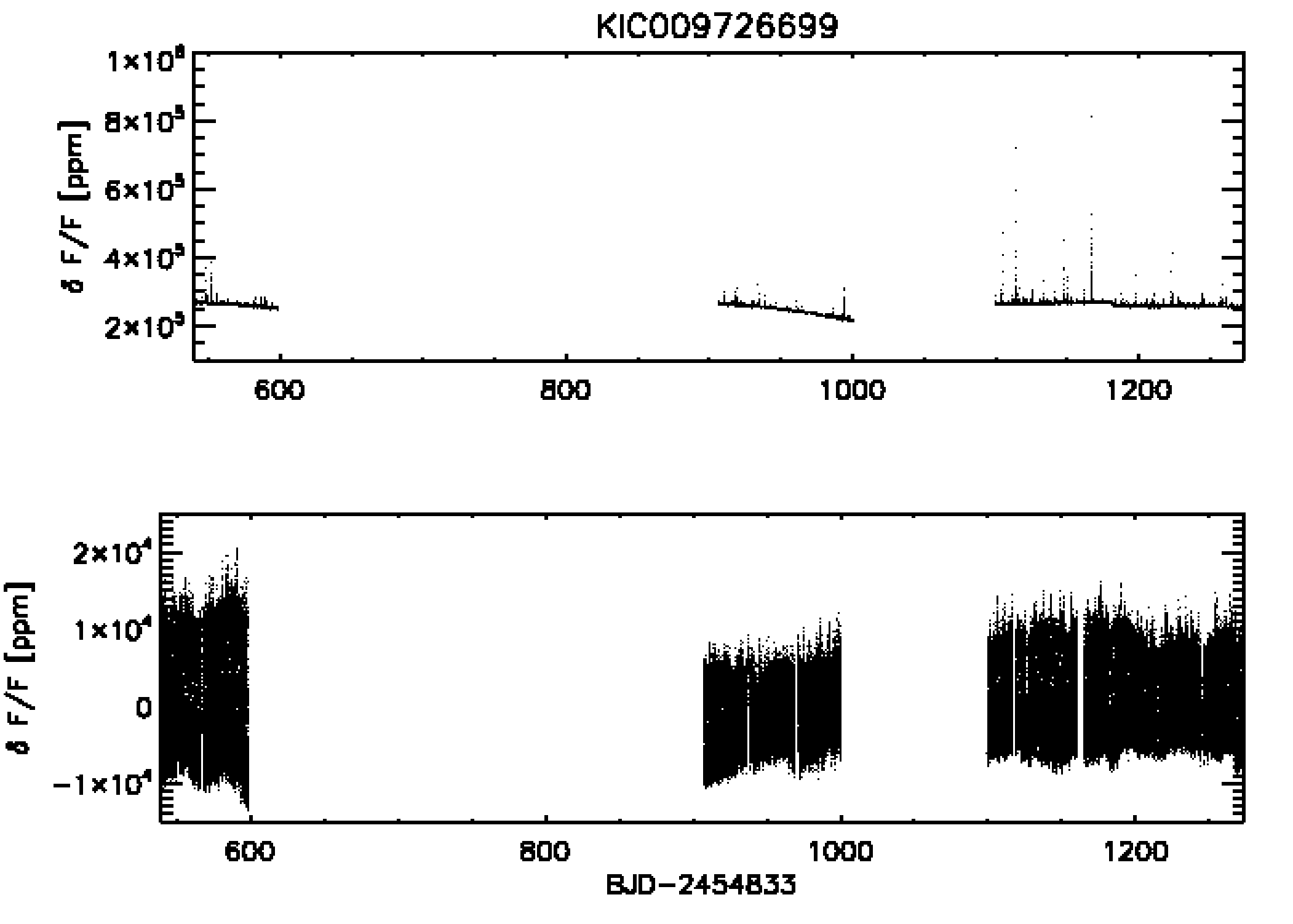}} &
   \resizebox{0.48\linewidth}{6cm}{\includegraphics{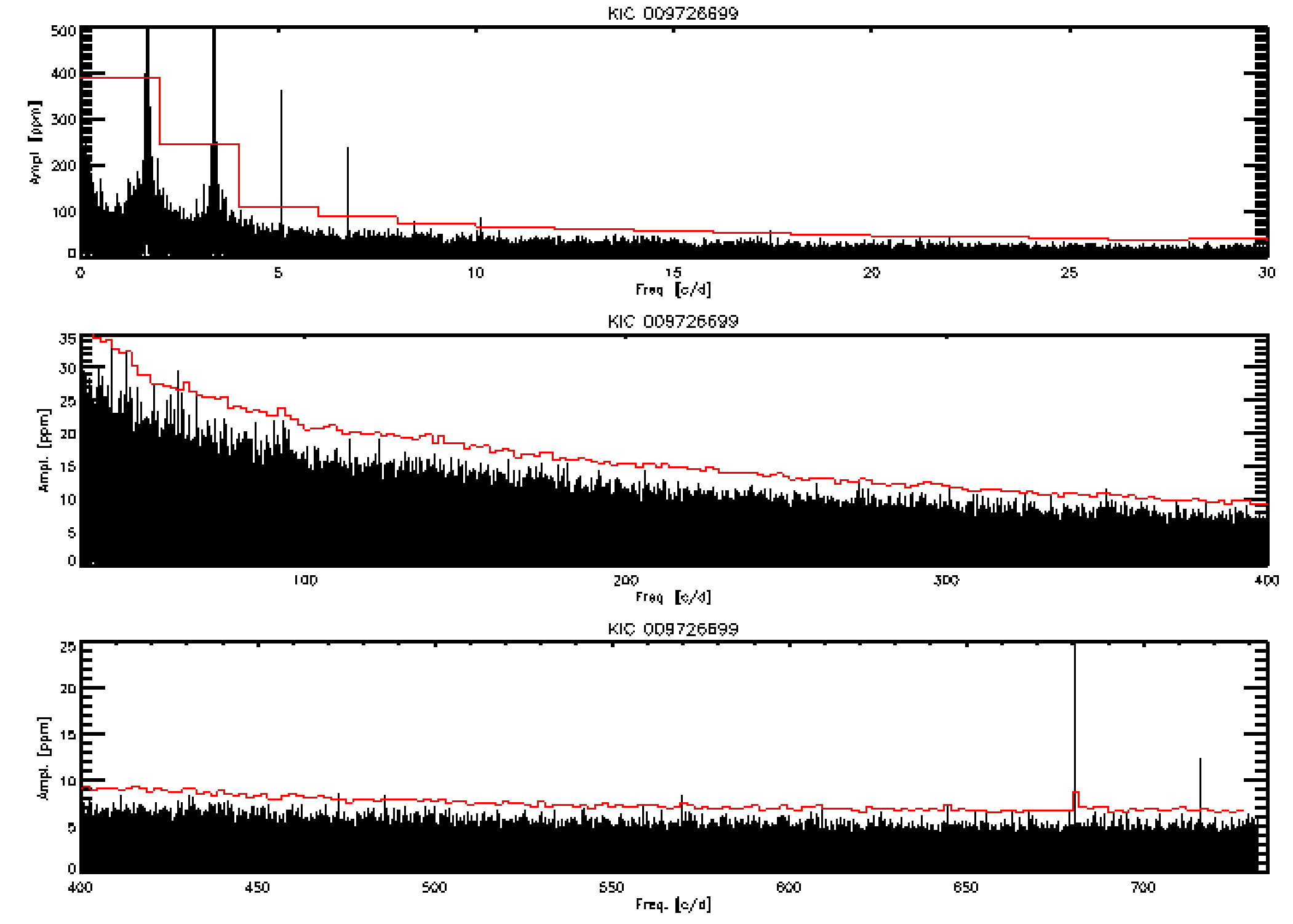}} \\
      \end{tabular}
   \caption{Left: KIC\,009726699 light curve quarters: Q6 (Q6.1 and Q6.2), Q10, Q12 and Q13. Above: original SAP light curves, note the high energy flares present in the data. Below: corrected for outliers with a 3$\sigma$ clipping of the two-point difference function, converted to relative flux in ppm and detrended with a 1 or 2 degree polynomial. Right: Amplitude spectrum -note the different abscissa ranges-. As the noise is so strongly dependent on the frequency, the detection threshold  is calculated in boxes of 2~c/d, and shown by the red line. See the text for more details.}
   \label{fig:9726699}
   \end{figure*}

  KIC\,009726699 was observed in Q10, Q12 and Q13, plus the two first months of Q6 public data were added to the analysis, which yields a time baseline of two years. This object is well documented in the literature as a very active M dwarf with very high energy flares. Because of the stable spots speckling its surface, its rotation period has been  determined to be 0.593~days by \cite{irwin11} and \cite{savanov11} from Q0 and Q1 long cadence data.

  The SAP light curves were corrected for flares and outliers and converted to relative flux in ppm; then a 1 or 2 degree polynomial was fitted. The original and detrended light curves are shown in Fig.~\ref{fig:9726699} (upper and lower left, respectively). The use of splines in this curve is tricky, as the star suffers very fast variations and the light curve is full of small and large flares. We found that the fitting of splines larger than 1~h was useless to completely remove the rotational modulation found in the light curve. Using 1~h splines, the rotational frequency and some of its harmonics were still recovered with amplitudes in the range of 50 to 100~ppm, so we decided to perform the FT analysis on the light curve without any spline correction. 

  The highest amplitude frequency is the well documented rotational frequency F1=1.687~c/d with an amplitude of 5025~ppm; additionally, its second to seventh harmonics are recovered in the periodogram and all of them are listed in Table~\ref{tab:kic97}. The strong rotational modulation produces a leakage into the frequencies closer to the rotation frequency, raising the detection threshold from about 80 to 400~ppm for frequencies lower than 10~c/d and progressively decaying until roughly a constant level between 9 to 7~ppm is reached in the 400 to 700~c/d range (see Fig.~\ref{fig:9726699} lower right). The amplitude spectrum is shown in Fig.~\ref{fig:9726699} (right): the upper plot is the low frequency range with the $y$-axis truncated by a factor of 10 to better see the harmonics of the main peak; the middle plot shows the frequency region between 30 and 400~c/d, while the bottom plot shows the frequency range up to the Nyquist frequency. The solid red line gives the detection threshold calculated in 2~c/d boxes, to highlight the high frequency dependence of noise.

\begin{table}
\caption{KIC\,009726699 frequencies}
\label{tab:kic97}
\centering
\renewcommand{\footnoterule}{}
\begin{tabular}{ccc}
\hline
  Freq.    &  Ampl.  & Comments   \\
  (c/d)    & (ppm)   &            \\
\hline
 1.687  &  5525  &  {\em f1} = f$_{rot}$   \\
 3.376  &  3385  &  {\em 2*f1} \\
 0.034  &  475   &   data download \\
 5.063  &  365   &  {\em 3*f1} \\
 6.753  &  240   &  {\em 4*f1} \\
10.125  &   90   &  {\em 6*f1} \\
 8.437  &   80   &  {\em 5*f1} \\
\end{tabular}
\end{table}

A frequency at 0.034~c/d, corresponding to a period of 29~days may be attributable to the monthly data downloads. No significant frequencies are found in the 30 to 400~c/d; we recall here that, even if most frequencies in the 30 to 100~c/d range lie above the solid red line in the middle plot of Fig.~\ref{fig:9726699}, this line is only a guidance, as the noise is evaluated locally in 2~c/d boxes around the frequency.

Finally, the 680.575 and 716.246~c/d frequencies stand out in the lower plot of Fig.~\ref{fig:9726699}; both, together with the 644.805~c/d frequency are identified with the so called 'U artifact' corresponding to combs of frequencies separated by 35.7~c/d that usually show the strongest peaks near the Nyquist frequency, as it is the case here. They can be matched to the U17b, U18b and U19b artifacts of table~1 in \cite{baran13b}.

None of the found frequencies could be attributed to pulsations.

\section{Conclusions}
  After Fourier Transform analysis of the four Kepler SC sample of M dwarfs -plus one M giant-, no convincing signal attributable to pulsations have been detected to a precision of a few parts per million. The significant frequencies detected (see Table~\ref{tab:spurious}  ) were identified with known artifacts of the short cadence Kepler data such as the harmonics of the long cadence sampling time or 'LC comb', the 'wide 20+', the U artifact and possibly the W peak.

  Some frequencies are found in the 320-445~c/d range (periods of 3 to 5~min) barely reaching the  4$\sigma$ detection limit. These frequencies are unstable as they change their amplitudes or even vanish during different months of the quarter with similar mean noises; therefore we considered them to be caused by the activity of the targets and not attributed to pulsations.

There are now ten M dwarfs -six from \cite{baran11a} and four in this work- searched for pulsation in high-cadence photometry reaching a detection limit of parts per million. Since this sample is still too small to be statistically significant, more short cadence Kepler observations of bright, quiet objects are highly desirable in pursuit of detection limits of parts per million, due to the expected low amplitude of the oscillations. These observations are the case of dedicated proposals to Kepler K2 mission\footnote{http://keplerscience.arc.nasa.gov/K2/} through the CARMENES Consortium\footnote{https://carmenes.caha.es/}. 

Our group also pursues the first detection of a pulsating M dwarf with high-cadence spectroscopy within the $\emph{Cool Tiny Beats Project}$ \citep{anglada2014, berdinas2014}, using HARPS \citep{mayor2003} and HARPS-N \citep{cosentino2014} high-resolution spectrographs to reach precisions of a few tens of cm/s in radial velocities (RV). Although there is not \emph{a priori} knowledge of the relation between photometric and RV amplitudes of the pulsation modes in M dwarfs, in the star FG Vir, the best studied $\delta$~Scuti-type object, the RV amplitudes in m/s are more than 70 times larger than the photometric amplitudes in mmag \citep{zima2006}. If it were to be the same for M dwarfs, a mode with an amplitude of 10~$\mu$mag would still produce a minimum RV amplitude of about 0.70~m/s, detectable in our on-going HARPS campaigns using the HARPS-Terra software \citep{anglada2012}. Moreover, it has been shown for the Solar data produced by the VIRGO \citep{virgo} and GOLF \citep{golf} experiments that the noise in the amplitude spectra caused by the variations produced by granulation in the photosphere of the star is much larger for the intensity measurements than for the radial velocities.

If pulsations were to be discovered, it would deeply impact our understanding of M dwarfs and all related science, such as planet and star formation and evolution, galactic evolution or constraints on the parameters of dark matter \citep{casanellas2013}. Discovery of pulsations will allow the application of asteroseismic methods to study M dwarf internal structure, providing, for instance, the mean density and the depth of the external convective layer. Most important is a method for determining the age of an M dwarf. The global properties of an M dwarf do not evolve significantly after they have reached the main sequence. However, pulsation spectra of M dwarfs have the potential to discriminate the age through regular period spacing of g-modes, which in our models, takes place even for modes near the fundamental radial mode. As an example, the period spacing between consecutive g-modes of the same degre $\ell$ for solar metalicity models of 0.60~M$\odot$ varies from about 45 to 25~min for models from 50 to 12\,000~Myr, respectively. This is a very powerful tool to achieve the grail of determining M dwarf ages. We just can hope that this exciting discovery of pulsations in M dwarfs is made soon.

\begin{table}
\caption{List of spurious frequencies found in the analysis, matched to the name given in    and the target they were found for (target names are abridged).}
\label{tab:spurious}
\centering
\renewcommand{\footnoterule}{}
\begin{tabular}{ccc}
\hline
  Freq. &  Artifact  & Object   \\
  (c/d) &            &            \\
\hline
 8.48   &  0.5*16.98 (spurious)  &  KIC\,24   \\
 19.92  &  wide 20+  &  KIC\,24 \\
 31.42  &  wide 20+  &  KIC\,24 \\
 31.68  &  wide 20+  &  KIC\,86 \\
 153.46 &   W?       &  KIC\,47 \\
 293.63 &   6/LC   &  KIC\,86 \\
 342.56 &   7/LC   &  KIC\,47, 86 \\
 391.54 &   8/LC   &  KIC\,24, 41, 86 \\
 440.48 &   9/LC   &  KIC\,24, 47, 86 \\
 644.81 &   U17b   &  KIC\,97 \\
 680.58 &   U18b   &  KIC\,97 \\
 716.25 &   U19b   &  KIC\,97 \\
\end{tabular}
\end{table}

\section*{Acknowledgments}

This paper includes data collected by the Kepler mission. Funding for the Kepler mission is provided by the NASA Science Mission Directorate. CR-L has a post-doctoral contract of the JAE-Doc program ``Junta para la ampliaci\'on de Estudios`` (CSIC) co-funded by the FSE (European Social Fund). CR-L and PJA acknowledge the funding provided by projects AYA2011-30147-C03-01 of the Spanish MINECO and 2011 FQM 7363 of Junta de Andaluc\'\i a. CR-L thanks R. Garc\'\i a and E. Rodr\'\i guez for useful discussion on the data. JG and JM acknowledge support from NASA under award No. NNX12AC90G. The authors thank the referee A. Baran for his suggestions that helped improved this paper.



\bibliographystyle{aa}
\bibliography{biblio}


\bsp

\label{lastpage}

\end{document}